\documentclass[12pt]{article}
\usepackage{times}
\usepackage{geometry}
\usepackage[utf8]{inputenc}
\usepackage{enumitem,amssymb}
\usepackage{ragged2e}
\usepackage[compact,small]{titlesec}
\usepackage[small,it]{caption}
\usepackage[pdftex]{graphicx}
\newlist{thematic}{itemize}{8}
\setlist[thematic]{label=$\square$}
\usepackage{pifont}
\newcommand{\cmark}{\ding{51}}%
\newcommand{\done}{\rlap{$\square$}{\raisebox{2pt}{\large\hspace{1pt}\cmark}}%
\hspace{-2.5pt}}

\usepackage[numbers,sort&compress]{natbib}
\usepackage{amsmath}
\usepackage{amssymb}

\newcommand{\vsvsmall}{\vspace*{-2pt}}

\begin{document}
\raggedright
\huge
Astro2020 Science White Paper \linebreak

Opportunities for Astrophysical Science from the Inner and Outer Solar System \linebreak
\normalsize

\noindent \textbf{Thematic Areas:} \hspace*{60pt} $\done$ Planetary Systems \hspace*{10pt} $\done$ Star and Planet Formation \hspace*{20pt}\linebreak
$\square$ Formation and Evolution of Compact Objects \hspace*{31pt} $\done$ Cosmology and Fundamental Physics \linebreak
  $\square$  Stars and Stellar Evolution \hspace*{1pt} $\square$ Resolved Stellar Populations and their Environments \hspace*{40pt} \linebreak
  $\done$    Galaxy Evolution   \hspace*{45pt} $\done$             Multi-Messenger Astronomy and Astrophysics \hspace*{65pt} \linebreak
  
\textbf{Principal Author:}

Name: Michael Zemcov	
 \linebreak						
Institution: The Rochester Institute of Technology \& the Jet Propulsion Laboratory
 \linebreak
Email: zemcov@cfd.rit.edu
 \linebreak
Phone: 585 475 2338 
 \linebreak
 
\textbf{Co-authors:} Iair Arcavi (Tel Aviv University), Richard G. Arendt (CRESST II/UMaryland/GSFC), Etienne Bachelet (Las Cumbres Observatory), Chas Beichman (JPL), James Bock (Caltech/JPL), Pontus Brandt (JHU-APL), Ranga Ram Chary (IPAC/Caltech), Asantha Cooray (UCI), Diana Dragomir (MIT), Varoujan Gorjian (JPL), Chester E.~Harman (NASA GISS), Richard Conn Henry (JHU), Carey Lisse (JHU-APL), Philip Lubin (UCSB), Shuji Matsuura (Kwansei Gakuin University), Ralph McNutt (JHU-APL), Jayant Murthy (Indian Institute of Astrophysics), Andrew R.~Poppe (UC Berkeley-SSL), Michael V.~Paul (JHU-APL), William T.~Reach (USRA/SOFIA), Yossi Shvartzvald (IPAC/Caltech), R.~A.~Street (Las Cumbres Observatory), Teresa Symons (RIT), Michael Werner (JPL)
  \linebreak

\textbf{Abstract:} Astrophysical measurements away from the 1 AU orbit of Earth can enable several astrophysical science cases that are challenging or impossible to perform from Earthbound platforms, including: building a detailed understanding of the extragalactic background light throughout the electromagnetic spectrum; measurements of the properties of dust and ice in the inner and outer solar system;  determinations of the mass of planets and stellar remnants far from luminous stars using gravitational microlensing; and stable time-domain astronomy.   Though potentially transformative for astrophysics, opportunities to fly instrumentation capable of these measurements are rare, and a mission to the distant solar system that includes instrumentation expressly designed to perform astrophysical science, or even one primarily for a different purpose but capable of precise astronomical investigation, has not yet been flown.  %Though such a mission is a costly and difficult endeavor requiring strong, committed collaboration between different space science disciplines, the potential scientific dividends are truly unique. 
In this White Paper, we describe the science motivations for this kind of measurement, and advocate for future flight opportunities that permit intersectional collaboration and cooperation to make these science investigations a reality.

\pagebreak

\section{Context}
\vsvsmall

The outer solar system is a unique, quiet vantage point from which to observe the universe around us.  At most wavelengths, the sensitivity of an instrument near the Earth 
is limited by light from the circumsolar dust cloud.  Reductions in this bright foreground would permit tremendous gains in sensitivity and temporal stability.  
However, we have been slow to take advantage of this resource.  Since Pioneer 10, there have been a relative handful of astrophysical studies using data from beyond the Earth's orbit \cite{Hanner1974, Toller1983, Holberg1985, Holberg1986,  Toller1987, Murthy1991, Murthy1993, Gordon1998, Murthy1999, Edelstein2000, Murthy2001, Matsuoka2011, Muraki2011, Gladstone2013, Zemcov2017, Matsumoto2018}, corresponding to a meager 3.5 results per decade. \textbf{In this White Paper we make the case that astrophysical observation well beyond the Earth's orbit can enable a wide range of virtually untapped astrophysical science.}

%These include the decrease in the light from interplanetary dust with heliocentric distance \cite{Hanner1974, Matsumoto2018}, Ly-$\alpha$ emission from the interplanetary medium \cite{Gladstone2013}, the diffuse light from the galaxy \cite{Toller1987, Gordon1998}, the brightness of the cosmic optical background \cite{Toller1983, Matsuoka2011, Zemcov2017} and the cosmic UV background \cite{Holberg1986, Murthy1991, Murthy1999, Edelstein2000}, exoplanet mass determination using gravitational lensing \cite{Muraki2011}, and the UV emission from specific objects \cite{Holberg1985} including studies of their spectral features \cite{Murthy1993, Murthy2001}.  In total, this corresponds to a meager 3.5 results per decade. 

%A future mission to the ISM offers an outstanding opportunity to perform astrophysical observations throughout and well beyond the solar system.  Using new technologies and detectors, an astrophysics instrument for an ISM probe could be made extremely compact and lightweight.  The science cases that are unique for such an instrument require only a small, 10 cm-class aperture and passively cooled, off-the-shelf detectors.  For studies of the dust disk of our solar system, a low spatial resolution FIR camera that shares the telescope focal plane would also be desirable.  Recent studies of a CubeSat-class astrophysical mission to the asteroid belt provide a good template for a workable low size, weight and power system. 

\section{Opening Novel Astrophysics From a Unique Vantage Point}
\vsvsmall

\subsection{Understanding the Solar Dust Cloud}  
\vsvsmall

Both the composition and structure of our circumsolar dust cloud is relatively well understood locally to the Earth \cite[\textit{e.g.}~\hspace{-4pt}][]{Leinert1998,Kelsall1998,MRR2013,Tsumura2013}.  Instruments on solar orbiting spacecraft such as \textit{Spitzer} have helped by providing Zodiacal Light (ZL) measurements along alternate lines of sight that are not constrained to originate at the Earth, and have highlighted the presence of local density enhancements in the ZL dust cloud at 1 AU \cite{Krick2012}.  However, beyond 1 AU we have little understanding of the structure of the interplanetary dust (IPD) cloud.  This is a major hindrance as we begin to probe the equivalent structures in exoplanetary systems \cite[\textit{e.g.}~review by][]{Hughes2018}.  Models indicate that there should be structures associated with the Edgeworth-Kuiper Belt (EKB; \cite{Poppe2016}), to which we see many analogs in the circumstellar disks around other stars.  We have virtually no understanding of how these disks map to our own, where we can hope to study composition and small-scale structure directly.  \textbf{Observations probing the light from IPD at a variety of wavelengths along different sight lines are necessary to develop a three-dimensional understanding of the morphology of our own dust disk and to contrast it with those of exoplanetary systems}.

\begin{figure*}[t]
\centering
\includegraphics*[width=6.5in]{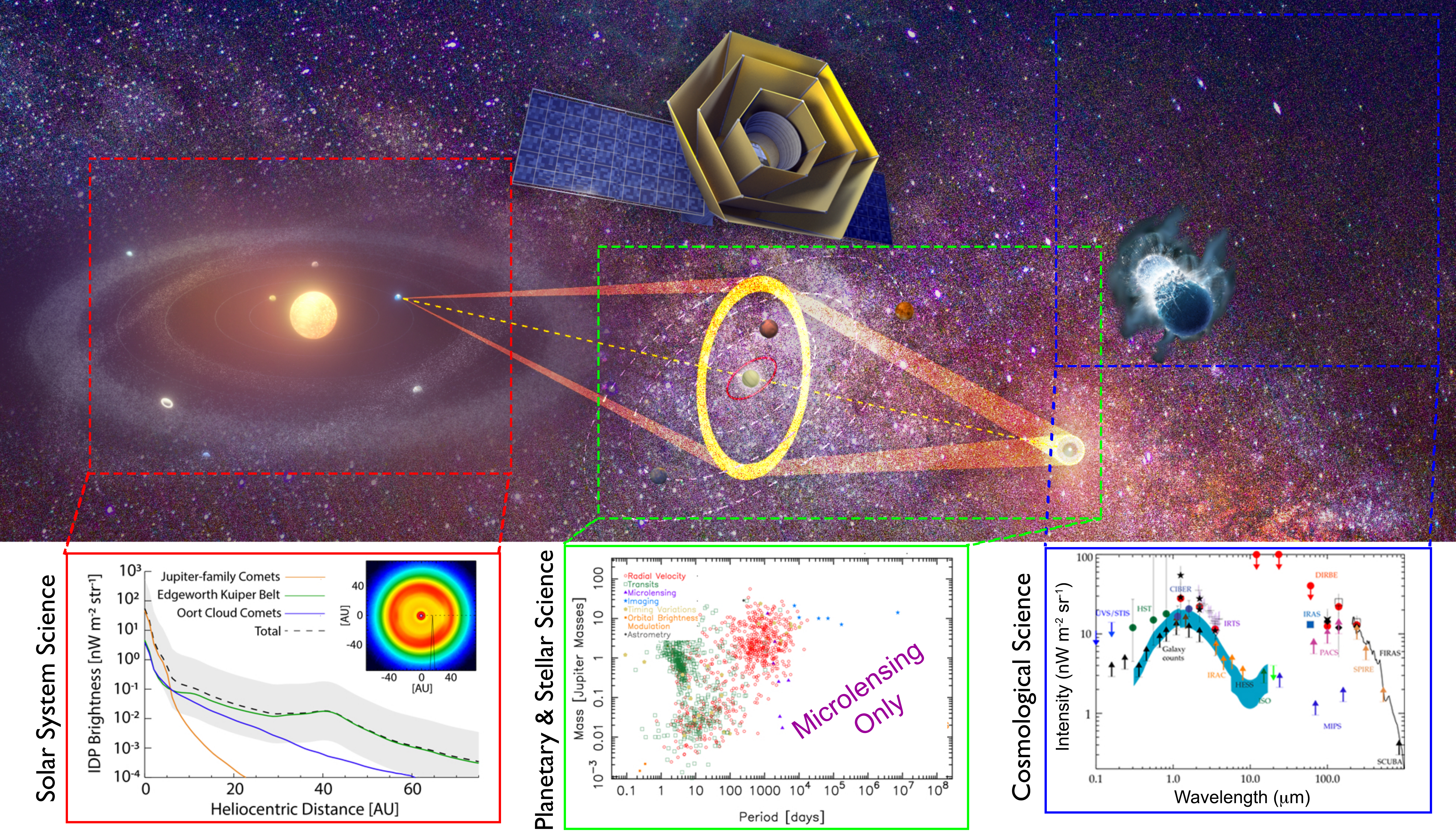}
\caption{Observations from other vantage points in the solar system touch on a wide range of astrophysics. \vspace{-25pt}}
\rule{\textwidth}{0.5pt}\vspace{-10pt}
\end{figure*}

\paragraph{Tomography of the Dust Cloud}  
%The zodiacal light (ZL) and the interplanetary dust (IPD) cloud itself are
%subjects of great interest. Any single observation of the ZL is an integration along the line of sight (LOS) through the IPD cloud, plus the
%emission of the Galaxy and the extragalactic background (EBL). The DIRBE data set provides ZL measurements of the the widest range of IR wavelengths and solar elongations, though not for a complete year. 
Even with observations taken in many directions and over the course of a full year, converting measurements from 1 AU into a precise model of the IPD cloud structure and optical properties is challenging.  By comparing data from instruments in different solar orbits, we can %make %integrated
observe from different perspectives, compiling different combinations of data that probe different aspects of the spatial distribution of IPD.
%and difference measurements made simultaneously along the co-linear line of sight, and as the relative separations of the instruments changed, the sampled portion of the IPD cloud would naturally vary. %Relatively short separations would provide the most localized sampling of the IPD cloud. This could prove very advantageous for  discerning small scale structure in the IPD clouds such as cometary dust trails. 
Orbits at $\lesssim 1 \,$AU could provide a rapidly varying sampling of the interior portion of the IDP cloud, while measurements at $>  1 \,$AU %will necessarily have relatively slow changes in the chords that are sampled, but 
could sample many more lines of sight through the cloud.  Such tomographic measurements would permit a three-dimensional map of the IPD cloud to be developed, which would address %, which relates to the production of IDPs by comets, asteroids, and KBOs, and the gravitational interaction of IDP orbits with the major planets.  %This would be particularly interesting in the outer solar system, where we observe large enhancements in the reflected light in exoplanetary systems.  
a variety of scientific questions, including the nature, composition, and evolution of dust outgassing from EKB objects and \"{O}ort-cloud comets, whether our models for dust transport in planetary systems are correct, and how our solar system's structure relates to that we see around other stars.

%Does dust outgassed by \"{O}ort-cloud comets as they enter the solar system contribute to the diffuse optical/NIR emission at large heliocentric distances?  When do Oort cloud comets begin outgassing as they enter the inner Solar System, and what is the composition of their dust?  Do our models for dust transport reproduce the observations?  How are the EKB structures emplaced, and how does our solar system relate to observations of exoplanetary systems?  These are all important questions for which we have no data.

\paragraph{Distant Look-Back Observations} In the next decade we will begin to find and characterize Earth-like planets orbiting other stars, and one of the most important validations available is the suite of observations that simulate the Earth as an extrasolar planet, including direct observations \cite[\textit{e.g.}~\hspace{-4pt}][]{Schreier2018}, Earthshine \cite[\textit{e.g.}~\hspace{-4pt}][]{Woolf2002}, instrument calibration \cite[\textit{e.g.}~\hspace{-4pt}][]{Robinson2011, Crow2011}, spacecraft flybys \cite{Sagan1993, Christensen1997}, and publicity shots from 40 AU \cite{Sagan1997}. 
%The Interstellar Probe mission could provide this much-needed and unique dataset for the Earth-like exoplanets that we expect to discover and characterize using the next generation of ground- and space-based telescopes (e.g., Batalha et al., 2015; France et al., 2015; Currie et al., 2018; Fortney et al., 2018; Gaudi et al., 2018).
Viewing the Earth from a large distance is the best analog to exoplanet observations \cite[\textit{e.g.}~\hspace{-4pt}][]{Roberge2012}, and allows for the validation of both forward models and model retrievals. Adding large-separation spectrographic data would represent the only ground-truthed, close-to interstellar observations of a habitable -- and inhabited -- Earth-like planet. 
%Lastly, the long duration of the proposed mission would allow not only snapshots, but observations taken at particular times using the same instrument, providing clear measurements of seasonality and the influence of viewing geometry in the spectra. The Earth need not be resolved to make these observations, which saves having to fly an ~11-meter telescope to 1,000 AU.

\subsection{Extragalactic Backgrounds}
\vsvsmall

The Extragalactic Background Light (EBL) is the cumulative sum of all radiation released over cosmic time, including light from galaxies throughout cosmic history, as well as any truly diffuse extragalactic sources \cite{Hauser2001,Cooray2016}. Measurements of the EBL can constrain galaxy formation and the evolution of cosmic structure, provide unique constraints on the Epoch of Reionization, and allow searches for beyond-standard model physics \cite{Tyson1995}.  %The EBL therefore allows a stringent cosmic consistency test wherein the observed brightness can strongly constrain structure formation models and simulations.  
The absolute brightness of the EBL has been established from Earth at many radio and X-ray wavelengths, but at most infrared, optical, and UV wavelengths a precise assessment of the sky brightness has been hampered by reflected and emitted light from IPD, which results in an irreducible $>50\,$\% uncertainty (and at some wavelengths significantly larger) on the absolute emission from the EBL  \cite[\textit{e.g.}~][]{Hauser1998}.  \textbf{By observing beyond the interplanetary dust, observations from the outer solar system can eliminate these uncertainties and definitively determine the absolute brightness of the EBL.}

\paragraph{The Optical/Near-IR EBL}  The optical/near-IR EBL encodes direct emission from stars integrated over time, so constrains the aggregate stellar population of the universe and nucleosynthesis in stars through cosmic history.  %Based on measurements near Earth, the optical/near-IR is not known to within a factor of three. This is primarily because of the uncertainty associated with subtracting the sunlight scattered off the local zodiacal dust cloud which is at least ten times brighter than the EBL contribution. 
By measuring the intensity and spectrum of the diffuse optical/near-IR EBL between 0.3 and 10 microns, we can: perform a census of the total mass density in stars and the fraction of them in diffuse structures; search for sources of diffuse emission which might arise from dark matter annihilation; determine the fraction of baryons that have been processed through stars and active galactic nuclei during the epoch of reionization; and understand the rate at which stars and supermassive black holes build up over cosmic time. %the contribution of the EoR may, in most scenarios, rise above the zodiacal light in the minimum around 4um where the scattered sunlight and the thermal reemission intersect.  Thus a careful set of measurements and component analysis of the type alluded to above should produce the first direct evaluation of the EoR.

% got to here
\paragraph{The Mid-IR/Far-IR} %Despite pioneering work on source counts with \textit{Spitzer}, \textit{Herschel} and ground-based millimeter/submillimeter surveys, our census of the total EBL at mid- and far-IR wavelengths ($ > 10 \, \mu$m) is incomplete.   
Here the EBL is dominated by thermal emission from small dust grains in galaxies, with high redshift sources from cosmic noon making the largest contribution \cite{Lagache2005}. %The EBL longward of $200 \, \mu$m provides the strongest constraints on dust obscured star-formation at $z>3$. 
By measuring the far-IR EBL, we can reveal the contribution from low-mass star-forming galaxies and thereby obtain a complete census of obscured star formation, measure obscured AGN activity, and trace the growth of dust and its evolution as a function of metallicity and cosmic time. %This is particularly important since there is evidence, based on the spectra of distant QSOs, that low-metallicity dust has different extinction properties from dust in local star-forming regions (e.g. Maiolino et al. 2010). 
Ultimately, this measurement offers a way to trace the evolution of the stellar initial mass function over time, which is one of the key uncertain parameters required in the conversion of luminosity to baryonic matter density. 

\paragraph{The Ultraviolet}  In the UV, the diffuse astrophysical background is thought to be largely due to light from local O and B stars scattered from dust in the ISM. %While we do have excellent imaging of UV radiation near 1500 \AA, we are lacking the spectroscopic information that is vital to confirming the source of the radiation.  Isolating the extragalactic component that quantifies the ionizing intergalactic radiation field from the local galactic component is a challenge.  
Advanced spectral decomposition techniques are required to separate the extragalactic component from dust and atomic scattering, as well as other emission processes \cite[\textit{e.g.}~\hspace{-4pt}][]{Murthy2009}.  %There is tentative evidence (Henry, Murthy, Overduin 2018) from Voyager-mission UV spectroscopy that at shorter wavelengths ($\sim 1100 \, $\AA) the spectrum is non-stellar and may extend shortward of the Lyman limit, indicative of a source of the diffuse UV background not explained by stars in our galaxy.  
Spectroscopic measurements far from the scattered solar light will help elucidate the origin of the galactic and extragalactic UV background, including any exotic physics that may be present \cite{Henry2018}.

\subsection{Breaking Mass-Distance Degeneracies in Gravitational Microlensing}
\vsvsmall

Photometry of stars in our galaxy can detect microlensing of the galactic source population, and observations of the microlensing light curves obtained at two locations in the solar system allows us to break degeneracies in models for the masses and parallax of the lensing system \cite{Gould1992, Buchalter1997}.  Microlensing is the most effective method for finding exoplanets beyond the snow line of their stars, where the sensitivity of other planet discovery techniques drops off rapidly, with 53 systems detected so far \cite{Perryman2018}.  \textbf{Because this method does not require the detection of light from the lens itself, it allows the detection and weighing of not only free-floating planets and brown dwarfs \cite{Mroz2017}, but also compact stellar remnants like black holes \cite{Wyrzykowski2011}.}

As has been demonstrated using observations from the Earth and \textit{Spitzer}, \textit{Kepler}, and \textit{EPOXI} \cite{Dong2007, Yee2015, Street2016, Zhu2017a}, stellar and planetary mass lensing can be observed with suitable facilities far from Earth.  A given lensing event will project into some radius in the solar system, which is characteristic of the mass of the lens and its geometry.  As an example, a 1 $M_{\odot}$ object at 4 kpc lensing a source at 8 kpc can be viewed within a $r= 4.0 \,$AU region wherein observers at different positions in the solar system will see the object lens the source star with different maxima and times of peak magnification.  An observatory further out in the solar system is sensitive to much larger masses, with a $10 \, M_{\odot}$ black hole at 4 kpc lensing a source at 8 kpc having a characteristic radius of $25 \,$AU.  Surveying towards the Bulge, Plane and Magellanic Clouds would enable us to explore the populations of low-mass stellar and planetary systems along multiple lines of sight and hence give insight into the distribution of these objects in different evolutionary environments in the Galaxy, and offers the unique possibility of measuring the mass function of quiescent black holes of $M \sim 20 \,$M$_{\odot}$ to allow us to distinguish whether they originate from stellar evolution \cite{Elbert2018} or possibly in the early Universe \cite{Carr2016}. 

\subsection{Time Domain Astrophysics}
\vsvsmall

Though already many instruments take advantage of the quiet environment away from the Earth at the L2 point of the Earth-Sun system  (\textit{e.g.}~\textit{WMAP}, \textit{Herschel}, \textit{Planck}, \textit{Gaia}, with \textit{JWST}, \textit{Euclid}, and \textit{WFIRST} planned), larger physical separations and even smaller temporal variability can be achieved elsewhere in the solar system.  \textbf{A platform away from Earth offers the possibility of a uniquely quiet and stable environment from which to make observations.}

\paragraph{Hyper-Stable Photometry} Any observation requiring stability on long time scales (exoplanet detection, SN light curves, variable star photometry) would benefit from access to the outer solar system.  The stable thermal and RF environment would permit instruments that are not affected by slow annual variations that can be present \cite[\textit{e.g.}~\hspace{-4pt}][]{Bennett2012,PlanckVIII}. 

\paragraph{Transient Counterpart Indentification}  Astronomical transients, such as supernovae, kilonovae/macronovae (merging neutron stars), and tidal disruption events are important laboratories of extreme physics.  Critical phases of these events, or even entire events, can be missed due to an unfavorable geometry of the Earth, the Sun, and the event. One notable example is the recent counterpart to the gravitational wave event GW170817 \cite{LIGO170817, LIGOmma}; had GW170817 occurred just one week later, it would have been Sun-constrained to Earth-based ultraviolet, optical, and infrared telescopes, the electromagnetic counterpart would not have been found, and the broad insights gained from the event would have been lost. Similarly, the nearest superluminous supernova to date, SN 2017egm, became unobservable due to Sun constraints just $2{–}3$ weeks after peak brightness \cite[\textit{e.g.}~\hspace{-4pt}][]{Nicholl2017, Bose2018}. %Also, Type IIP supernovae have a light-curve plateau which lasts almost always 100 days - a time period which is difficult to observe continuously from Earth. The physical mechanisms responsible for powering superluminous supernovae and maintaining Type IIP plateaus are still debated. More events with coverage during the relevant phases would greatly inform emission models.
A platform elsewhere in the solar system could observe events that are Sun-constrained from Earth, and thus provide both unique observations at critical phases and wavelength coverage not available from ground-based platforms.

%An additional strength of any space based platform for transient science is the ability to observe in the ultraviolet. The critical early phases of supernovae and kilonovae, as well as most phases of tidal disruption events, emit mostly in the ultraviolet. This emission contains unique clues regarding the progenitors of supernovae, the emission mechanisms of neutron star mergers, and the physics of accretion onto supermassive black holes, respectively. 

\begin{table}[t]
    \caption{Summary of science cases and requirements.\vspace{-10pt}}
    \label{tab:my_label}
    \centering \small
    \begin{tabular}{|l|c|c|c|c|}
    \hline
    Science Topic & Type & Wavelength & Angular & Heliocentric \\ 
    & & Range & Resolution & Distance \\ \hline
Solar IPD Structure & Spectrographic Survey & Optical/Near-IR & $\sim 10^{\prime \prime}$ & $<10 \,$AU \\
EKB Disk & Spectrographic Survey & Far-IR & $\sim 10^{\prime \prime}$ & $>10 \,$AU \\
Earth Imaging & Pointed Spectro-photometry & Optical/Near-IR & $< 1^{\prime \prime}$ & $> 100 \,$AU \\
Absolute EBL & Spectrographic Survey & UV to Sub-mm & $\sim 10^{\prime \prime}$ & $> 5 \,$AU \\ 
Microlensing & Pointed Photometry & Optical/Near-IR & $< 1^{\prime \prime}$ & Any \\
Transient Follow-Up & Pointed Spectro-photometry & Optical/Near-IR & $\sim 1^{\prime \prime}$ & Any \\ \hline
%\vspace{-10pt}
    \end{tabular}
\end{table}

\section{Strawman Mission Concepts}

Though it could have a transformative impact on a wide range of astrophysical fields \cite[\textit{e.g.~\hspace{-4pt}}][]{Zemcov2018}, a mission to the outer solar system that includes instrumentation expressly designed to perform astrophysical science has not yet been flown.  Previous proposals for both stand-alone missions \cite{Mather1996, Matsuura2014} or astronomical instruments piggybacked on other missions  \cite{Bock2012} have proven more politically than technically challenging.  This is unfortunate, as a piggyback concept is a cost-effective way to multiply the science return of expensive missions to the outer solar system.  \textbf{Strong advocacy from the astrophysical community could make positive collaboration and cooperation between the different NASA divisions a  realistic outcome.}

A possibility that has been discussed over the years is an Interstellar Probe to the pristine ISM \cite{Holzer1991,Liewer2000,McNutt2001,Mewaldt2001,Fiehler2006,Wimmer2009,Stone2015}.  The current incarnation of this concept would travel to 1000 AU in a 50-year mission \cite{McNutt2017}.  Astrophysical measurements during its cruise phase would offer a unique opportunity to generate both high-impact science during the long quiescent periods en route to the ISM, as well as to build and maintain technical expertise in the spacecraft and instruments over the generations of scientists and engineers required to execute such a mission.  An Interstellar Probe could be a true flagship of space science, offering an unique opportunity to fulfill some of the promise of astrophysical observation far from Earth.

% References
\bibliographystyle{unsrtnat}
\bibliography{oss_decadal}  

\begin{thebibliography}{67}
\providecommand{\natexlab}[1]{#1}
\providecommand{\url}[1]{\texttt{#1}}
\expandafter\ifx\csname urlstyle\endcsname\relax
  \providecommand{\doi}[1]{doi: #1}\else
  \providecommand{\doi}{doi: \begingroup \urlstyle{rm}\Url}\fi

\bibitem[{Hanner} et~al.(1974){Hanner}, {Weinberg}, {DeShields}, {Green}, and
  {Toller}]{Hanner1974}
M.~S. {Hanner}, J.~L. {Weinberg}, L.~M. {DeShields}, II, B.~A. {Green}, and
  G.~N. {Toller}.
\newblock {Zodiacal light and the asteroid belt: The view from Pioneer 10}.
\newblock \emph{\jgr}, 79:\penalty0 3671, 1974.
\newblock \doi{10.1029/JA079i025p03671}.

\bibitem[{Toller}(1983)]{Toller1983}
G.~N. {Toller}.
\newblock {The extragalactic background light at 4400 \AA}.
\newblock \emph{\apjl}, 266:\penalty0 L79--L82, March 1983.
\newblock \doi{10.1086/183982}.

\bibitem[{Holberg} and {Barber}(1985)]{Holberg1985}
J.~B. {Holberg} and H.~B. {Barber}.
\newblock {Far-ultraviolet background observations at high galactic latitude. I
  - The Coma Cluster}.
\newblock \emph{\apj}, 292:\penalty0 16--21, May 1985.
\newblock \doi{10.1086/163128}.

\bibitem[{Holberg}(1986)]{Holberg1986}
J.~B. {Holberg}.
\newblock {Far-ultraviolet background observations at high galactic latitude.
  II - Diffuse emission}.
\newblock \emph{\apj}, 311:\penalty0 969--978, December 1986.
\newblock \doi{10.1086/164834}.

\bibitem[{Toller} et~al.(1987){Toller}, {Tanabe}, and {Weinberg}]{Toller1987}
G.~{Toller}, H.~{Tanabe}, and J.~L. {Weinberg}.
\newblock {Background starlight at the north and south celestial, ecliptic, and
  galactic poles}.
\newblock \emph{\aap}, 188:\penalty0 24--34, December 1987.

\bibitem[{Murthy} et~al.(1991){Murthy}, {Henry}, and {Holberg}]{Murthy1991}
J.~{Murthy}, R.~C. {Henry}, and J.~B. {Holberg}.
\newblock {Constraints on the optical properties of interstellar dust in the
  far-ultraviolet - Voyager observations of the diffuse sky background}.
\newblock \emph{\apj}, 383:\penalty0 198--204, December 1991.
\newblock \doi{10.1086/170776}.

\bibitem[{Murthy} et~al.(1993){Murthy}, {Im}, {Henry}, and
  {Holberg}]{Murthy1993}
J.~{Murthy}, M.~{Im}, R.~C. {Henry}, and J.~B. {Holberg}.
\newblock {Voyager Observations of Diffuse Far-Ultraviolet Continuum and Line
  Emission in Eridanus}.
\newblock \emph{\apj}, 419:\penalty0 739, December 1993.
\newblock \doi{10.1086/173524}.

\bibitem[{Gordon} et~al.(1998){Gordon}, {Witt}, and {Friedmann}]{Gordon1998}
K.~D. {Gordon}, A.~N. {Witt}, and B.~C. {Friedmann}.
\newblock {Detection of Extended Red Emission in the Diffuse Interstellar
  Medium}.
\newblock \emph{\apj}, 498:\penalty0 522--540, May 1998.
\newblock \doi{10.1086/305571}.

\bibitem[{Murthy} et~al.(1999){Murthy}, {Hall}, {Earl}, {Henry}, and
  {Holberg}]{Murthy1999}
J.~{Murthy}, D.~{Hall}, M.~{Earl}, R.~C. {Henry}, and J.~B. {Holberg}.
\newblock {An Analysis of 17 Years of Voyager Observations of the Diffuse
  Far-Ultraviolet Radiation Field}.
\newblock \emph{\apj}, 522:\penalty0 904--914, September 1999.
\newblock \doi{10.1086/307652}.

\bibitem[{Edelstein} et~al.(2000){Edelstein}, {Bowyer}, and
  {Lampton}]{Edelstein2000}
J.~{Edelstein}, S.~{Bowyer}, and M.~{Lampton}.
\newblock {Reanalysis ofVoyager Ultraviolet Spectrometer Limits to the
  Extreme-Ultraviolet and Far-Ultraviolet Diffuse Astronomical Flux}.
\newblock \emph{\apj}, 539:\penalty0 187--190, August 2000.
\newblock \doi{10.1086/309192}.

\bibitem[{Murthy} et~al.(2001){Murthy}, {Henry}, {Shelton}, and
  {Holberg}]{Murthy2001}
J.~{Murthy}, R.~C. {Henry}, R.~L. {Shelton}, and J.~B. {Holberg}.
\newblock {Upper Limits on O VI Emission From Voyager Observations}.
\newblock \emph{\apjl}, 557:\penalty0 L47--L50, August 2001.
\newblock \doi{10.1086/323041}.

\bibitem[{Matsuoka} et~al.(2011){Matsuoka}, {Ienaka}, {Kawara}, and
  {Oyabu}]{Matsuoka2011}
Y.~{Matsuoka}, N.~{Ienaka}, K.~{Kawara}, and S.~{Oyabu}.
\newblock {Cosmic Optical Background: The View from Pioneer 10/11}.
\newblock \emph{\apj}, 736:\penalty0 119, August 2011.
\newblock \doi{10.1088/0004-637X/736/2/119}.

\bibitem[{Muraki} et~al.(2011){Muraki}, {Han}, {Bennett}, {Suzuki}, {Monard},
  {Street}, {Jorgensen}, {Kundurthy}, {Skowron}, {Becker}, {Albrow},
  {Fouqu{\'e}}, {Heyrovsk{\'y}}, {Barry}, {Beaulieu}, {Wellnitz}, {Bond},
  {Sumi}, {Dong}, {Gaudi}, {Bramich}, {Dominik}, {Abe}, {Botzler}, {Freeman},
  {Fukui}, {Furusawa}, {Hayashi}, {Hearnshaw}, {Hosaka}, {Itow}, {Kamiya},
  {Korpela}, {Kilmartin}, {Lin}, {Ling}, {Makita}, {Masuda}, {Matsubara},
  {Miyake}, {Nishimoto}, {Ohnishi}, {Perrott}, {Rattenbury}, {Saito},
  {Skuljan}, {Sullivan}, {Sweatman}, {Tristram}, {Wada}, {Yock}, {MOA
  Collaboration}, {Christie}, {DePoy}, {Gorbikov}, {Gould}, {Kaspi}, {Lee},
  {Mallia}, {Maoz}, {McCormick}, {Moorhouse}, {Natusch}, {Park}, {Pogge},
  {Polishook}, {Shporer}, {Thornley}, {Yee}, {{$\mu$}FUN Collaboration},
  {Allan}, {Browne}, {Horne}, {Kains}, {Snodgrass}, {Steele}, {Tsapras},
  {RoboNet Collaboration}, {Batista}, {Bennett}, {Brillant}, {Caldwell},
  {Cassan}, {Cole}, {Corrales}, {Coutures}, {Dieters}, {Dominis Prester},
  {Donatowicz}, {Greenhill}, {Kubas}, {Marquette}, {Martin}, {Menzies}, {Sahu},
  {Waldman}, {Williams}, {Zub}, {PLANET Collaboration}, {Bourhrous},
  {Matsuoka}, {Nagayama}, {Oi}, {Randriamanakoto}, {IRSF Observers}, {Bozza},
  {Burgdorf}, {Calchi Novati}, {Dreizler}, {Finet}, {Glitrup}, {Harps{\o}e},
  {Hinse}, {Hundertmark}, {Liebig}, {Maier}, {Mancini}, {Mathiasen}, {Rahvar},
  {Ricci}, {Scarpetta}, {Skottfelt}, {Surdej}, {Southworth}, {Wambsganss},
  {Zimmer}, {MiNDSTEp Consortium}, {Udalski}, {Poleski}, {Wyrzykowski},
  {Ulaczyk}, {Szyma{\'n}ski}, {Kubiak}, {Pietrzy{\'n}ski}, {Soszy{\'n}ski}, and
  {OGLE Collaboration}]{Muraki2011}
Y.~{Muraki}, C.~{Han}, D.~P. {Bennett}, D.~{Suzuki}, L.~A.~G. {Monard},
  R.~{Street}, U.~G. {Jorgensen}, P.~{Kundurthy}, J.~{Skowron}, A.~C. {Becker},
  M.~D. {Albrow}, P.~{Fouqu{\'e}}, D.~{Heyrovsk{\'y}}, R.~K. {Barry}, J.-P.
  {Beaulieu}, D.~D. {Wellnitz}, I.~A. {Bond}, T.~{Sumi}, S.~{Dong}, B.~S.
  {Gaudi}, D.~M. {Bramich}, M.~{Dominik}, F.~{Abe}, C.~S. {Botzler},
  M.~{Freeman}, A.~{Fukui}, K.~{Furusawa}, F.~{Hayashi}, J.~B. {Hearnshaw},
  S.~{Hosaka}, Y.~{Itow}, K.~{Kamiya}, A.~V. {Korpela}, P.~M. {Kilmartin},
  W.~{Lin}, C.~H. {Ling}, S.~{Makita}, K.~{Masuda}, Y.~{Matsubara},
  N.~{Miyake}, K.~{Nishimoto}, K.~{Ohnishi}, Y.~C. {Perrott}, N.~J.
  {Rattenbury}, T.~{Saito}, L.~{Skuljan}, D.~J. {Sullivan}, W.~L. {Sweatman},
  P.~J. {Tristram}, K.~{Wada}, P.~C.~M. {Yock}, {MOA Collaboration}, G.~W.
  {Christie}, D.~L. {DePoy}, E.~{Gorbikov}, A.~{Gould}, S.~{Kaspi}, C.-U.
  {Lee}, F.~{Mallia}, D.~{Maoz}, J.~{McCormick}, D.~{Moorhouse}, T.~{Natusch},
  B.-G. {Park}, R.~W. {Pogge}, D.~{Polishook}, A.~{Shporer}, G.~{Thornley},
  J.~C. {Yee}, {{$\mu$}FUN Collaboration}, A.~{Allan}, P.~{Browne}, K.~{Horne},
  N.~{Kains}, C.~{Snodgrass}, I.~{Steele}, Y.~{Tsapras}, {RoboNet
  Collaboration}, V.~{Batista}, C.~S. {Bennett}, S.~{Brillant}, J.~A.~R.
  {Caldwell}, A.~{Cassan}, A.~{Cole}, R.~{Corrales}, C.~{Coutures},
  S.~{Dieters}, D.~{Dominis Prester}, J.~{Donatowicz}, J.~{Greenhill},
  D.~{Kubas}, J.-B. {Marquette}, R.~{Martin}, J.~{Menzies}, K.~C. {Sahu},
  I.~{Waldman}, A.~{Williams}, M.~{Zub}, {PLANET Collaboration},
  H.~{Bourhrous}, Y.~{Matsuoka}, T.~{Nagayama}, N.~{Oi}, Z.~{Randriamanakoto},
  {IRSF Observers}, V.~{Bozza}, M.~J. {Burgdorf}, S.~{Calchi Novati},
  S.~{Dreizler}, F.~{Finet}, M.~{Glitrup}, K.~{Harps{\o}e}, T.~C. {Hinse},
  M.~{Hundertmark}, C.~{Liebig}, G.~{Maier}, L.~{Mancini}, M.~{Mathiasen},
  S.~{Rahvar}, D.~{Ricci}, G.~{Scarpetta}, J.~{Skottfelt}, J.~{Surdej},
  J.~{Southworth}, J.~{Wambsganss}, F.~{Zimmer}, {MiNDSTEp Consortium},
  A.~{Udalski}, R.~{Poleski}, {\L}.~{Wyrzykowski}, K.~{Ulaczyk}, M.~K.
  {Szyma{\'n}ski}, M.~{Kubiak}, G.~{Pietrzy{\'n}ski}, I.~{Soszy{\'n}ski}, and
  {OGLE Collaboration}.
\newblock {Discovery and Mass Measurements of a Cold, 10 Earth Mass Planet and
  Its Host Star}.
\newblock \emph{\apj}, 741:\penalty0 22, November 2011.
\newblock \doi{10.1088/0004-637X/741/1/22}.

\bibitem[{Gladstone} et~al.(2013){Gladstone}, {Stern}, and
  {Pryor}]{Gladstone2013}
G.~R. {Gladstone}, S.~A. {Stern}, and W.~R. {Pryor}.
\newblock \emph{{New Horizons Cruise Observations of Lyman-{$\alpha$} Emissions
  from the Interplanetary Medium}}, page 177.
\newblock 2013.
\newblock \doi{10.1007/978-1-4614-6384-9\_6}.

\bibitem[{Zemcov} et~al.(2017){Zemcov}, {Immel}, {Nguyen}, {Cooray}, {Lisse},
  and {Poppe}]{Zemcov2017}
M.~{Zemcov}, P.~{Immel}, C.~{Nguyen}, A.~{Cooray}, C.~M. {Lisse}, and A.~R.
  {Poppe}.
\newblock {Measurement of the cosmic optical background using the long range
  reconnaissance imager on New Horizons}.
\newblock \emph{Nature Communications}, 8:\penalty0 15003, April 2017.
\newblock \doi{10.1038/ncomms15003}.

\bibitem[{Matsumoto} et~al.(2018){Matsumoto}, {Tsumura}, {Matsuoka}, and
  {Pyo}]{Matsumoto2018}
T.~{Matsumoto}, K.~{Tsumura}, Y.~{Matsuoka}, and J.~{Pyo}.
\newblock {Zodiacal Light Beyond Earth Orbit Observed with Pioneer 10}.
\newblock \emph{\aj}, 156:\penalty0 86, September 2018.
\newblock \doi{10.3847/1538-3881/aad0f0}.

\bibitem[{Leinert} et~al.(1998){Leinert}, {Bowyer}, {Haikala}, {Hanner},
  {Hauser}, {Levasseur-Regourd}, {Mann}, {Mattila}, {Reach}, {Schlosser},
  {Staude}, {Toller}, {Weiland}, {Weinberg}, and {Witt}]{Leinert1998}
C.~{Leinert}, S.~{Bowyer}, L.~K. {Haikala}, M.~S. {Hanner}, M.~G. {Hauser},
  A.-C. {Levasseur-Regourd}, I.~{Mann}, K.~{Mattila}, W.~T. {Reach},
  W.~{Schlosser}, H.~J. {Staude}, G.~N. {Toller}, J.~L. {Weiland}, J.~L.
  {Weinberg}, and A.~N. {Witt}.
\newblock {The 1997 reference of diffuse night sky brightness}.
\newblock \emph{\aaps}, 127:\penalty0 1--99, January 1998.
\newblock \doi{10.1051/aas:1998105}.

\bibitem[{Kelsall} et~al.(1998){Kelsall}, {Weiland}, {Franz}, {Reach},
  {Arendt}, {Dwek}, {Freudenreich}, {Hauser}, {Moseley}, {Odegard},
  {Silverberg}, and {Wright}]{Kelsall1998}
T.~{Kelsall}, J.~L. {Weiland}, B.~A. {Franz}, W.~T. {Reach}, R.~G. {Arendt},
  E.~{Dwek}, H.~T. {Freudenreich}, M.~G. {Hauser}, S.~H. {Moseley}, N.~P.
  {Odegard}, R.~F. {Silverberg}, and E.~L. {Wright}.
\newblock {The COBE Diffuse Infrared Background Experiment Search for the
  Cosmic Infrared Background. II. Model of the Interplanetary Dust Cloud}.
\newblock \emph{\apj}, 508:\penalty0 44--73, November 1998.
\newblock \doi{10.1086/306380}.

\bibitem[{Rowan-Robinson} and {May}(2013)]{MRR2013}
M.~{Rowan-Robinson} and B.~{May}.
\newblock {An improved model for the infrared emission from the zodiacal dust
  cloud: cometary, asteroidal and interstellar dust}.
\newblock \emph{\mnras}, 429:\penalty0 2894--2902, March 2013.
\newblock \doi{10.1093/mnras/sts471}.

\bibitem[{Tsumura} et~al.(2013){Tsumura}, {Matsumoto}, {Matsuura}, {Sakon}, and
  {Wada}]{Tsumura2013}
K.~{Tsumura}, T.~{Matsumoto}, S.~{Matsuura}, I.~{Sakon}, and T.~{Wada}.
\newblock {Low-Resolution Spectrum of the Extragalactic Background Light with
  the AKARI InfraRed Camera}.
\newblock \emph{\pasj}, 65:\penalty0 121, December 2013.
\newblock \doi{10.1093/pasj/65.6.121}.

\bibitem[{Krick} et~al.(2012){Krick}, {Glaccum}, {Carey}, {Lowrance}, {Surace},
  {Ingalls}, {Hora}, and {Reach}]{Krick2012}
J.~E. {Krick}, W.~J. {Glaccum}, S.~J. {Carey}, P.~J. {Lowrance}, J.~A.
  {Surace}, J.~G. {Ingalls}, J.~L. {Hora}, and W.~T. {Reach}.
\newblock {A Spitzer/IRAC Measure of the Zodiacal Light}.
\newblock \emph{\apj}, 754:\penalty0 53, July 2012.
\newblock \doi{10.1088/0004-637X/754/1/53}.

\bibitem[{Hughes} et~al.(2018){Hughes}, {Duch{\^e}ne}, and
  {Matthews}]{Hughes2018}
A.~M. {Hughes}, G.~{Duch{\^e}ne}, and B.~C. {Matthews}.
\newblock {Debris Disks: Structure, Composition, and Variability}.
\newblock \emph{\araa}, 56:\penalty0 541--591, September 2018.
\newblock \doi{10.1146/annurev-astro-081817-052035}.

\bibitem[{Poppe}(2016)]{Poppe2016}
A.~R. {Poppe}.
\newblock {An improved model for interplanetary dust fluxes in the outer Solar
  System}.
\newblock \emph{\icarus}, 264:\penalty0 369--386, January 2016.
\newblock \doi{10.1016/j.icarus.2015.10.001}.

\bibitem[{Schreier} et~al.(2018){Schreier}, {St{\"a}dt}, {Hedelt}, and
  {Godolt}]{Schreier2018}
F.~{Schreier}, S.~{St{\"a}dt}, P.~{Hedelt}, and M.~{Godolt}.
\newblock {Transmission spectroscopy with the ACE-FTS infrared spectral atlas
  of Earth: A model validation and feasibility study}.
\newblock \emph{Molecular Astrophysics}, 11:\penalty0 1--22, June 2018.
\newblock \doi{10.1016/j.molap.2018.02.001}.

\bibitem[{Woolf} et~al.(2002){Woolf}, {Smith}, {Traub}, and {Jucks}]{Woolf2002}
N.~J. {Woolf}, P.~S. {Smith}, W.~A. {Traub}, and K.~W. {Jucks}.
\newblock {The Spectrum of Earthshine: A Pale Blue Dot Observed from the
  Ground}.
\newblock \emph{\apj}, 574:\penalty0 430--433, July 2002.
\newblock \doi{10.1086/340929}.

\bibitem[{Robinson} et~al.(2011){Robinson}, {Meadows}, {Crisp}, {Deming},
  {A'Hearn}, {Charbonneau}, {Livengood}, {Seager}, {Barry}, {Hearty},
  {Hewagama}, {Lisse}, {McFadden}, and {Wellnitz}]{Robinson2011}
T.~D. {Robinson}, V.~S. {Meadows}, D.~{Crisp}, D.~{Deming}, M.~F. {A'Hearn},
  D.~{Charbonneau}, T.~A. {Livengood}, S.~{Seager}, R.~K. {Barry}, T.~{Hearty},
  T.~{Hewagama}, C.~M. {Lisse}, L.~A. {McFadden}, and D.~D. {Wellnitz}.
\newblock {Earth as an Extrasolar Planet: Earth Model Validation Using EPOXI
  Earth Observations}.
\newblock \emph{Astrobiology}, 11:\penalty0 393--408, June 2011.
\newblock \doi{10.1089/ast.2011.0642}.

\bibitem[{Crow} et~al.(2011){Crow}, {McFadden}, {Robinson}, {Meadows},
  {Livengood}, {Hewagama}, {Barry}, {Deming}, {Lisse}, and
  {Wellnitz}]{Crow2011}
C.~A. {Crow}, L.~A. {McFadden}, T.~{Robinson}, V.~S. {Meadows}, T.~A.
  {Livengood}, T.~{Hewagama}, R.~K. {Barry}, L.~D. {Deming}, C.~M. {Lisse}, and
  D.~{Wellnitz}.
\newblock {Views from EPOXI: Colors in Our Solar System as an Analog for
  Extrasolar Planets}.
\newblock \emph{\apj}, 729:\penalty0 130, March 2011.
\newblock \doi{10.1088/0004-637X/729/2/130}.

\bibitem[{Sagan} et~al.(1993){Sagan}, {Thompson}, {Carlson}, {Gurnett}, and
  {Hord}]{Sagan1993}
C.~{Sagan}, W.~R. {Thompson}, R.~{Carlson}, D.~{Gurnett}, and C.~{Hord}.
\newblock {A search for life on Earth from the Galileo spacecraft}.
\newblock \emph{\nat}, 365:\penalty0 715--721, October 1993.
\newblock \doi{10.1038/365715a0}.

\bibitem[{Christensen} and {Pearl}(1997)]{Christensen1997}
P.~R. {Christensen} and J.~C. {Pearl}.
\newblock {Initial data from the Mars Global Surveyor thermal emission
  spectrometer experiment: Observations of the Earth}.
\newblock \emph{\jgr}, 102:\penalty0 10875--10880, May 1997.
\newblock \doi{10.1029/97JE00637}.

\bibitem[Sagan(1997)]{Sagan1997}
C.~Sagan.
\newblock \emph{Billions and Billions: Thoughts on Life and Death at the Brink
  of the Millennium}.
\newblock Random House, 1997.
\newblock ISBN 9780679411604.
\newblock URL \url{https://books.google.com/books?id=tsbaAAAAMAAJ}.

\bibitem[{Roberge} et~al.(2012){Roberge}, {Chen}, {Millan-Gabet}, {Weinberger},
  {Hinz}, {Stapelfeldt}, {Absil}, {Kuchner}, and {Bryden}]{Roberge2012}
A.~{Roberge}, C.~H. {Chen}, R.~{Millan-Gabet}, A.~J. {Weinberger}, P.~M.
  {Hinz}, K.~R. {Stapelfeldt}, O.~{Absil}, M.~J. {Kuchner}, and G.~{Bryden}.
\newblock {The Exozodiacal Dust Problem for Direct Observations of Exo-Earths}.
\newblock \emph{\pasp}, 124:\penalty0 799, August 2012.
\newblock \doi{10.1086/667218}.

\bibitem[{Hauser} and {Dwek}(2001)]{Hauser2001}
M.~G. {Hauser} and E.~{Dwek}.
\newblock {The Cosmic Infrared Background: Measurements and Implications}.
\newblock \emph{\araa}, 39:\penalty0 249--307, 2001.
\newblock \doi{10.1146/annurev.astro.39.1.249}.

\bibitem[{Cooray}(2016)]{Cooray2016}
A.~{Cooray}.
\newblock {Extragalactic background light measurements and applications}.
\newblock \emph{Royal Society Open Science}, 3\penalty0 (15):\penalty0 150555,
  March 2016.
\newblock \doi{10.1098/rsos.150555}.

\bibitem[{Tyson}(1995)]{Tyson1995}
J.~A. {Tyson}.
\newblock {The optical extragalactic background radiation.}
\newblock In D.~{Calzetti}, M.~{Livio}, and P.~{Madau}, editors, \emph{{\it
  Extragalactic Background Radiation Meeting} (Ed. Calzetti, D. et al.)
  103-133}, pages 103--133, 1995.

\bibitem[{Hauser} et~al.(1998){Hauser}, {Arendt}, {Kelsall}, {Dwek}, {Odegard},
  {Weiland}, {Freudenreich}, {Reach}, {Silverberg}, {Moseley}, {Pei}, {Lubin},
  {Mather}, {Shafer}, {Smoot}, {Weiss}, {Wilkinson}, and {Wright}]{Hauser1998}
M.~G. {Hauser}, R.~G. {Arendt}, T.~{Kelsall}, E.~{Dwek}, N.~{Odegard}, J.~L.
  {Weiland}, H.~T. {Freudenreich}, W.~T. {Reach}, R.~F. {Silverberg}, S.~H.
  {Moseley}, Y.~C. {Pei}, P.~{Lubin}, J.~C. {Mather}, R.~A. {Shafer}, G.~F.
  {Smoot}, R.~{Weiss}, D.~T. {Wilkinson}, and E.~L. {Wright}.
\newblock {The COBE Diffuse Infrared Background Experiment Search for the
  Cosmic Infrared Background. I. Limits and Detections}.
\newblock \emph{\apj}, 508:\penalty0 25--43, November 1998.
\newblock \doi{10.1086/306379}.

\bibitem[{Lagache} et~al.(2005){Lagache}, {Puget}, and {Dole}]{Lagache2005}
G.~{Lagache}, J.-L. {Puget}, and H.~{Dole}.
\newblock {Dusty Infrared Galaxies: Sources of the Cosmic Infrared Background}.
\newblock \emph{\araa}, 43:\penalty0 727--768, September 2005.
\newblock \doi{10.1146/annurev.astro.43.072103.150606}.

\bibitem[{Murthy}(2009)]{Murthy2009}
J.~{Murthy}.
\newblock {Observations of the near and far ultraviolet background}.
\newblock \emph{\apss}, 320:\penalty0 21--26, April 2009.
\newblock \doi{10.1007/s10509-008-9855-y}.

\bibitem[{Conn Henry} et~al.(2018){Conn Henry}, {Murthy}, and
  {Overduin}]{Henry2018}
Richard {Conn Henry}, Jayant {Murthy}, and James {Overduin}.
\newblock {Discovery of an Ionizing Radiation Field in the Universe}.
\newblock \emph{arXiv e-prints}, art. arXiv:1805.09658, May 2018.

\bibitem[{Gould}(1992)]{Gould1992}
A.~{Gould}.
\newblock {Extending the MACHO search to about 10 exp 6 solar masses}.
\newblock \emph{\apj}, 392:\penalty0 442--451, June 1992.
\newblock \doi{10.1086/171443}.

\bibitem[{Buchalter} and {Kamionkowski}(1997)]{Buchalter1997}
A.~{Buchalter} and M.~{Kamionkowski}.
\newblock {Rates for Parallax-shifted Microlensing Events from Ground-based
  Observations of the Galactic Bulge}.
\newblock \emph{\apj}, 482:\penalty0 782--791, June 1997.
\newblock \doi{10.1086/304163}.

\bibitem[Perryman(2018)]{Perryman2018}
M.~Perryman.
\newblock \emph{The Exoplanet Handbook}.
\newblock Cambridge University Press, 2018.
\newblock ISBN 9781108419772.
\newblock URL \url{https://books.google.com/books?id=ngtmDwAAQBAJ}.

\bibitem[{Mroz} et~al.(2017){Mroz}, {Ryu}, {Skowron}, {Udalski}, {Gould},
  {Szymanski}, {Soszynski}, {Poleski}, {Pietrukowicz}, {Kozlowski}, {Pawlak},
  {Ulaczyk}, {Albrow}, {Chung}, {Jung}, {Han}, {Hwang}, {Shin}, {Yee}, {Zhu},
  {Cha}, {Kim}, {Kim}, {Kim}, {Lee}, {Lee}, {Lee}, {Park}, and
  {Pogge}]{Mroz2017}
P.~{Mroz}, Y.-H. {Ryu}, J.~{Skowron}, A.~{Udalski}, A.~{Gould}, M.~K.
  {Szymanski}, I.~{Soszynski}, R.~{Poleski}, P.~{Pietrukowicz}, S.~{Kozlowski},
  M.~{Pawlak}, K.~{Ulaczyk}, M.~D. {Albrow}, S.-J. {Chung}, Y.~K. {Jung},
  C.~{Han}, K.-H. {Hwang}, I.-G. {Shin}, J.~C. {Yee}, W.~{Zhu}, S.-M. {Cha},
  D.-J. {Kim}, H.-W. {Kim}, S.-L. {Kim}, C.-U. {Lee}, D.-J. {Lee}, Y.~{Lee},
  B.-G. {Park}, and R.~W. {Pogge}.
\newblock {A free-floating planet candidate from the OGLE and KMTNet surveys}.
\newblock \emph{ArXiv e-prints}, December 2017.

\bibitem[{Wyrzykowski} et~al.(2011){Wyrzykowski}, {Skowron}, {Koz{\l}owski},
  {Udalski}, {Szyma{\'n}ski}, {Kubiak}, {Pietrzy{\'n}ski}, {Soszy{\'n}ski},
  {Szewczyk}, {Ulaczyk}, {Poleski}, and {Tisserand}]{Wyrzykowski2011}
L.~{Wyrzykowski}, J.~{Skowron}, S.~{Koz{\l}owski}, A.~{Udalski}, M.~K.
  {Szyma{\'n}ski}, M.~{Kubiak}, G.~{Pietrzy{\'n}ski}, I.~{Soszy{\'n}ski},
  O.~{Szewczyk}, K.~{Ulaczyk}, R.~{Poleski}, and P.~{Tisserand}.
\newblock {The OGLE view of microlensing towards the Magellanic Clouds - IV.
  OGLE-III SMC data and final conclusions on MACHOs}.
\newblock \emph{\mnras}, 416:\penalty0 2949--2961, October 2011.
\newblock \doi{10.1111/j.1365-2966.2011.19243.x}.

\bibitem[{Dong} et~al.(2007){Dong}, {Udalski}, {Gould}, {Reach}, {Christie},
  {Boden}, {Bennett}, {Fazio}, {Griest}, {Szyma{\'n}ski}, {Kubiak},
  {Soszy{\'n}ski}, {Pietrzy{\'n}ski}, {Szewczyk}, {Wyrzykowski}, {Ulaczyk},
  {Wieckowski}, {Paczy{\'n}ski}, {DePoy}, {Pogge}, {Preston}, {Thompson}, and
  {Patten}]{Dong2007}
S.~{Dong}, A.~{Udalski}, A.~{Gould}, W.~T. {Reach}, G.~W. {Christie}, A.~F.
  {Boden}, D.~P. {Bennett}, G.~{Fazio}, K.~{Griest}, M.~K. {Szyma{\'n}ski},
  M.~{Kubiak}, I.~{Soszy{\'n}ski}, G.~{Pietrzy{\'n}ski}, O.~{Szewczyk},
  {\L}.~{Wyrzykowski}, K.~{Ulaczyk}, T.~{Wieckowski}, B.~{Paczy{\'n}ski}, D.~L.
  {DePoy}, R.~W. {Pogge}, G.~W. {Preston}, I.~B. {Thompson}, and B.~M.
  {Patten}.
\newblock {First Space-Based Microlens Parallax Measurement: Spitzer
  Observations of OGLE-2005-SMC-001}.
\newblock \emph{\apj}, 664:\penalty0 862--878, August 2007.
\newblock \doi{10.1086/518536}.

\bibitem[{Yee} et~al.(2015){Yee}, {Gould}, {Beichman}, {Calchi Novati},
  {Carey}, {Gaudi}, {Henderson}, {Nataf}, {Penny}, {Shvartzvald}, and
  {Zhu}]{Yee2015}
J.~C. {Yee}, A.~{Gould}, C.~{Beichman}, S.~{Calchi Novati}, S.~{Carey}, B.~S.
  {Gaudi}, C.~B. {Henderson}, D.~{Nataf}, M.~{Penny}, Y.~{Shvartzvald}, and
  W.~{Zhu}.
\newblock {Criteria for Sample Selection to Maximize Planet Sensitivity and
  Yield from Space-Based Microlens Parallax Surveys}.
\newblock \emph{\apj}, 810:\penalty0 155, September 2015.
\newblock \doi{10.1088/0004-637X/810/2/155}.

\bibitem[{Street} et~al.(2016){Street}, {Udalski}, {Calchi Novati},
  {Hundertmark}, {Zhu}, {Gould}, {Yee}, {Tsapras}, {Bennett}, {RoboNet
  Project}, {Consortium}, {J{\o}rgensen}, {Dominik}, {Andersen}, {Bachelet},
  {Bozza}, {Bramich}, {Burgdorf}, {Cassan}, {Ciceri}, {D'Ago}, {Dong}, {Evans},
  {Gu}, {Harkonnen}, {Hinse}, {Horne}, {Figuera Jaimes}, {Kains}, {Kerins},
  {Korhonen}, {Kuffmeier}, {Mancini}, {Menzies}, {Mao}, {Peixinho}, {Popovas},
  {Rabus}, {Rahvar}, {Ranc}, {Tronsgaard Rasmussen}, {Scarpetta}, {Schmidt},
  {Skottfelt}, {Snodgrass}, {Southworth}, {Steele}, {Surdej}, {Unda-Sanzana},
  {Verma}, {von Essen}, {Wambsganss}, {Wang}, {Wertz}, {OGLE Project},
  {Poleski}, {Pawlak}, {Szyma{\'n}ski}, {Skowron}, {Mr{\'o}z}, {Koz{\l}owski},
  {Wyrzykowski}, {Pietrukowicz}, {Pietrzy{\'n}ski}, {Soszy{\'n}ski}, {Ulaczyk},
  {Spitzer Team}, {Beichman}, {Bryden}, {Carey}, {Gaudi}, {Henderson}, {Pogge},
  {Shvartzvald}, {MOA Collaboration}, {Abe}, {Asakura}, {Bhattacharya}, {Bond},
  {Donachie}, {Freeman}, {Fukui}, {Hirao}, {Inayama}, {Itow}, {Koshimoto},
  {Li}, {Ling}, {Masuda}, {Matsubara}, {Muraki}, {Nagakane}, {Nishioka},
  {Ohnishi}, {Oyokawa}, {Rattenbury}, {Saito}, {Sharan}, {Sullivan}, {Sumi},
  {Suzuki}, {Tristram}, {Wakiyama}, {Yonehara}, {KMTNet Modeling Team}, {Han},
  {Choi}, {Park}, {Jung}, and {Shin}]{Street2016}
R.~A. {Street}, A.~{Udalski}, S.~{Calchi Novati}, M.~P.~G. {Hundertmark},
  W.~{Zhu}, A.~{Gould}, J.~{Yee}, Y.~{Tsapras}, D.~P. {Bennett}, T.~{RoboNet
  Project}, M.~{Consortium}, U.~G. {J{\o}rgensen}, M.~{Dominik}, M.~I.
  {Andersen}, E.~{Bachelet}, V.~{Bozza}, D.~M. {Bramich}, M.~J. {Burgdorf},
  A.~{Cassan}, S.~{Ciceri}, G.~{D'Ago}, S.~{Dong}, D.~F. {Evans}, S.-h. {Gu},
  H.~{Harkonnen}, T.~C. {Hinse}, K.~{Horne}, R.~{Figuera Jaimes}, N.~{Kains},
  E.~{Kerins}, H.~{Korhonen}, M.~{Kuffmeier}, L.~{Mancini}, J.~{Menzies},
  S.~{Mao}, N.~{Peixinho}, A.~{Popovas}, M.~{Rabus}, S.~{Rahvar}, C.~{Ranc},
  R.~{Tronsgaard Rasmussen}, G.~{Scarpetta}, R.~{Schmidt}, J.~{Skottfelt},
  C.~{Snodgrass}, J.~{Southworth}, I.~A. {Steele}, J.~{Surdej},
  E.~{Unda-Sanzana}, P.~{Verma}, C.~{von Essen}, J.~{Wambsganss}, Y.-B. {Wang},
  O.~{Wertz}, T.~{OGLE Project}, R.~{Poleski}, M.~{Pawlak}, M.~K.
  {Szyma{\'n}ski}, J.~{Skowron}, P.~{Mr{\'o}z}, S.~{Koz{\l}owski},
  {\L}.~{Wyrzykowski}, P.~{Pietrukowicz}, G.~{Pietrzy{\'n}ski},
  I.~{Soszy{\'n}ski}, K.~{Ulaczyk}, {Spitzer Team}, C.~{Beichman}, G.~{Bryden},
  S.~{Carey}, B.~S. {Gaudi}, C.~B. {Henderson}, R.~W. {Pogge},
  Y.~{Shvartzvald}, {MOA Collaboration}, F.~{Abe}, Y.~{Asakura},
  A.~{Bhattacharya}, I.~A. {Bond}, M.~{Donachie}, M.~{Freeman}, A.~{Fukui},
  Y.~{Hirao}, K.~{Inayama}, Y.~{Itow}, N.~{Koshimoto}, M.~C.~A. {Li}, C.~H.
  {Ling}, K.~{Masuda}, Y.~{Matsubara}, Y.~{Muraki}, M.~{Nagakane},
  T.~{Nishioka}, K.~{Ohnishi}, H.~{Oyokawa}, N.~{Rattenbury}, T.~{Saito},
  A.~{Sharan}, D.~J. {Sullivan}, T.~{Sumi}, D.~{Suzuki}, J.~{Tristram},
  Y.~{Wakiyama}, A.~{Yonehara}, {KMTNet Modeling Team}, C.~{Han}, J.-Y. {Choi},
  H.~{Park}, Y.~K. {Jung}, and I.-G. {Shin}.
\newblock {Spitzer Parallax of OGLE-2015-BLG-0966: A Cold Neptune in the
  Galactic Disk}.
\newblock \emph{\apj}, 819:\penalty0 93, March 2016.
\newblock \doi{10.3847/0004-637X/819/2/93}.

\bibitem[{Zhu} et~al.(2017){Zhu}, {Udalski}, {Huang}, {Calchi Novati}, {Sumi},
  {Poleski}, {Skowron}, {Mr{\'o}z}, {Szyma{\'n}ski}, {Soszy{\'n}ski},
  {Pietrukowicz}, {Koz{\l}owski}, {Ulaczyk}, {Pawlak}, {OGLE Collaboration},
  {Beichman}, {Bryden}, {Carey}, {Gaudi}, {Gould}, {Henderson}, {Shvartzvald},
  {Yee}, {Spitzer Team}, {Bond}, {Bennett}, {Suzuki}, {Rattenbury},
  {Koshimoto}, {Abe}, {Asakura}, {Barry}, {Bhattacharya}, {Donachie}, {Evans},
  {Fukui}, {Hirao}, {Itow}, {Kawasaki}, {Li}, {Ling}, {Masuda}, {Matsubara},
  {Miyazaki}, {Munakata}, {Muraki}, {Nagakane}, {Ohnishi}, {Ranc}, {Saito},
  {Sharan}, {Sullivan}, {Tristram}, {Yamada}, {Yonehara}, and {MOA
  Collaboration}]{Zhu2017a}
W.~{Zhu}, A.~{Udalski}, C.~X. {Huang}, S.~{Calchi Novati}, T.~{Sumi},
  R.~{Poleski}, J.~{Skowron}, P.~{Mr{\'o}z}, M.~K. {Szyma{\'n}ski},
  I.~{Soszy{\'n}ski}, P.~{Pietrukowicz}, S.~{Koz{\l}owski}, K.~{Ulaczyk},
  M.~{Pawlak}, {OGLE Collaboration}, C.~{Beichman}, G.~{Bryden}, S.~{Carey},
  B.~S. {Gaudi}, A.~{Gould}, C.~B. {Henderson}, Y.~{Shvartzvald}, J.~C. {Yee},
  {Spitzer Team}, I.~A. {Bond}, D.~P. {Bennett}, D.~{Suzuki}, N.~J.
  {Rattenbury}, N.~{Koshimoto}, F.~{Abe}, Y.~{Asakura}, R.~K. {Barry},
  A.~{Bhattacharya}, M.~{Donachie}, P.~{Evans}, A.~{Fukui}, Y.~{Hirao},
  Y.~{Itow}, K.~{Kawasaki}, M.~C.~A. {Li}, C.~H. {Ling}, K.~{Masuda},
  Y.~{Matsubara}, S.~{Miyazaki}, H.~{Munakata}, Y.~{Muraki}, M.~{Nagakane},
  K.~{Ohnishi}, C.~{Ranc}, T.~{Saito}, A.~{Sharan}, D.~J. {Sullivan}, P.~J.
  {Tristram}, T.~{Yamada}, A.~{Yonehara}, and {MOA Collaboration}.
\newblock {An Isolated Microlens Observed from K2, Spitzer, and Earth}.
\newblock \emph{\apjl}, 849:\penalty0 L31, November 2017.
\newblock \doi{10.3847/2041-8213/aa93fa}.

\bibitem[{Elbert} et~al.(2018){Elbert}, {Bullock}, and
  {Kaplinghat}]{Elbert2018}
O.~D. {Elbert}, J.~S. {Bullock}, and M.~{Kaplinghat}.
\newblock {Counting black holes: The cosmic stellar remnant population and
  implications for LIGO}.
\newblock \emph{\mnras}, 473:\penalty0 1186--1194, January 2018.
\newblock \doi{10.1093/mnras/stx1959}.

\bibitem[{Carr} et~al.(2016){Carr}, {K{\"u}hnel}, and {Sandstad}]{Carr2016}
B.~{Carr}, F.~{K{\"u}hnel}, and M.~{Sandstad}.
\newblock {Primordial black holes as dark matter}.
\newblock \emph{\prd}, 94\penalty0 (8):\penalty0 083504, October 2016.
\newblock \doi{10.1103/PhysRevD.94.083504}.

\bibitem[{Bennett} et~al.(2013){Bennett}, {Larson}, {Weiland}, {Jarosik},
  {Hinshaw}, {Odegard}, {Smith}, {Hill}, {Gold}, {Halpern}, {Komatsu}, {Nolta},
  {Page}, {Spergel}, {Wollack}, {Dunkley}, {Kogut}, {Limon}, {Meyer}, {Tucker},
  and {Wright}]{Bennett2012}
C.~L. {Bennett}, D.~{Larson}, J.~L. {Weiland}, N.~{Jarosik}, G.~{Hinshaw},
  N.~{Odegard}, K.~M. {Smith}, R.~S. {Hill}, B.~{Gold}, M.~{Halpern},
  E.~{Komatsu}, M.~R. {Nolta}, L.~{Page}, D.~N. {Spergel}, E.~{Wollack},
  J.~{Dunkley}, A.~{Kogut}, M.~{Limon}, S.~S. {Meyer}, G.~S. {Tucker}, and
  E.~L. {Wright}.
\newblock {Nine-year Wilkinson Microwave Anisotropy Probe (WMAP) Observations:
  Final Maps and Results}.
\newblock \emph{\apjs}, 208:\penalty0 20, October 2013.
\newblock \doi{10.1088/0067-0049/208/2/20}.

\bibitem[{Planck Collaboration} et~al.(2014){Planck Collaboration}, {Ade},
  {Aghanim}, {Armitage-Caplan}, {Arnaud}, {Ashdown}, {Atrio-Barandela},
  {Aumont}, {Baccigalupi}, {Banday}, and et~al.]{PlanckVIII}
{Planck Collaboration}, P.~A.~R. {Ade}, N.~{Aghanim}, C.~{Armitage-Caplan},
  M.~{Arnaud}, M.~{Ashdown}, F.~{Atrio-Barandela}, J.~{Aumont},
  C.~{Baccigalupi}, A.~J. {Banday}, and et~al.
\newblock {Planck 2013 results. VIII. HFI photometric calibration and
  mapmaking}.
\newblock \emph{\aap}, 571:\penalty0 A8, November 2014.
\newblock \doi{10.1051/0004-6361/201321538}.

\bibitem[{Abbott} et~al.(2017{\natexlab{a}}){Abbott}, {Abbott}, {Abbott},
  {Acernese}, {Ackley}, {Adams}, {Adams}, {Addesso}, {Adhikari}, {Adya}, and
  et~al.]{LIGO170817}
B.~P. {Abbott}, R.~{Abbott}, T.~D. {Abbott}, F.~{Acernese}, K.~{Ackley},
  C.~{Adams}, T.~{Adams}, P.~{Addesso}, R.~X. {Adhikari}, V.~B. {Adya}, and
  et~al.
\newblock {GW170817: Observation of Gravitational Waves from a Binary Neutron
  Star Inspiral}.
\newblock \emph{Physical Review Letters}, 119\penalty0 (16):\penalty0 161101,
  October 2017{\natexlab{a}}.
\newblock \doi{10.1103/PhysRevLett.119.161101}.

\bibitem[{Abbott} et~al.(2017{\natexlab{b}}){Abbott}, {Abbott}, {Abbott},
  {Acernese}, {Ackley}, {Adams}, {Adams}, {Addesso}, {Adhikari}, {Adya}, and
  et~al.]{LIGOmma}
B.~P. {Abbott}, R.~{Abbott}, T.~D. {Abbott}, F.~{Acernese}, K.~{Ackley},
  C.~{Adams}, T.~{Adams}, P.~{Addesso}, R.~X. {Adhikari}, V.~B. {Adya}, and
  et~al.
\newblock {Multi-messenger Observations of a Binary Neutron Star Merger}.
\newblock \emph{\apjl}, 848:\penalty0 L12, October 2017{\natexlab{b}}.
\newblock \doi{10.3847/2041-8213/aa91c9}.

\bibitem[{Nicholl} et~al.(2017){Nicholl}, {Berger}, {Margutti}, {Blanchard},
  {Guillochon}, {Leja}, and {Chornock}]{Nicholl2017}
M.~{Nicholl}, E.~{Berger}, R.~{Margutti}, P.~K. {Blanchard}, J.~{Guillochon},
  J.~{Leja}, and R.~{Chornock}.
\newblock {The Superluminous Supernova SN 2017egm in the Nearby Galaxy NGC
  3191: A Metal-rich Environment Can Support a Typical SLSN Evolution}.
\newblock \emph{\apjl}, 845:\penalty0 L8, August 2017.
\newblock \doi{10.3847/2041-8213/aa82b1}.

\bibitem[{Bose} et~al.(2018){Bose}, {Dong}, {Pastorello}, {Filippenko},
  {Kochanek}, {Mauerhan}, {Romero-Ca{\~n}izales}, {Brink}, {Chen}, {Prieto},
  {Post}, {Ashall}, {Grupe}, {Tomasella}, {Benetti}, {Shappee}, {Stanek},
  {Cai}, {Falco}, {Lundqvist}, {Mattila}, {Mutel}, {Ochner}, {Pooley},
  {Stritzinger}, {Villanueva}, {Zheng}, {Beswick}, {Brown}, {Cappellaro},
  {Davis}, {Fraser}, {de Jaeger}, {Elias-Rosa}, {Gall}, {Gaudi}, {Herczeg},
  {Hestenes}, {Holoien}, {Hosseinzadeh}, {Hsiao}, {Hu}, {Jaejin}, {Jeffers},
  {Koff}, {Kumar}, {Kurtenkov}, {Lau}, {Prentice}, {Reynolds}, {Rudy},
  {Shahbandeh}, {Somero}, {Stassun}, {Thompson}, {Valenti}, {Woo}, and
  {Yunus}]{Bose2018}
S.~{Bose}, S.~{Dong}, A.~{Pastorello}, A.~V. {Filippenko}, C.~S. {Kochanek},
  J.~{Mauerhan}, C.~{Romero-Ca{\~n}izales}, T.~G. {Brink}, P.~{Chen}, J.~L.
  {Prieto}, R.~{Post}, C.~{Ashall}, D.~{Grupe}, L.~{Tomasella}, S.~{Benetti},
  B.~J. {Shappee}, K.~Z. {Stanek}, Z.~{Cai}, E.~{Falco}, P.~{Lundqvist},
  S.~{Mattila}, R.~{Mutel}, P.~{Ochner}, D.~{Pooley}, M.~D. {Stritzinger},
  S.~{Villanueva}, Jr., W.~{Zheng}, R.~J. {Beswick}, P.~J. {Brown},
  E.~{Cappellaro}, S.~{Davis}, M.~{Fraser}, T.~{de Jaeger}, N.~{Elias-Rosa},
  C.~{Gall}, B.~S. {Gaudi}, G.~J. {Herczeg}, J.~{Hestenes}, T.~W.-S. {Holoien},
  G.~{Hosseinzadeh}, E.~Y. {Hsiao}, S.~{Hu}, S.~{Jaejin}, B.~{Jeffers}, R.~A.
  {Koff}, S.~{Kumar}, A.~{Kurtenkov}, M.~W. {Lau}, S.~{Prentice},
  T.~{Reynolds}, R.~J. {Rudy}, M.~{Shahbandeh}, A.~{Somero}, K.~G. {Stassun},
  T.~A. {Thompson}, S.~{Valenti}, J.-H. {Woo}, and S.~{Yunus}.
\newblock {Gaia17biu/SN 2017egm in NGC 3191: The Closest Hydrogen-poor
  Superluminous Supernova to Date Is in a Normal, Massive, Metal-rich Spiral
  Galaxy}.
\newblock \emph{\apj}, 853:\penalty0 57, January 2018.
\newblock \doi{10.3847/1538-4357/aaa298}.

\bibitem[{Zemcov} et~al.(2018){Zemcov}, {Arcavi}, {Arendt}, {Bachelet}, {Ram
  Chary}, {Cooray}, {Dragomir}, {Conn Henry}, {Lisse}, {Matsuura}, {Murthy},
  {Nguyen}, {Poppe}, {Street}, and {Werner}]{Zemcov2018}
M.~{Zemcov}, I.~{Arcavi}, R.~{Arendt}, E.~{Bachelet}, R.~{Ram Chary},
  A.~{Cooray}, D.~{Dragomir}, R.~{Conn Henry}, C.~{Lisse}, S.~{Matsuura},
  J.~{Murthy}, C.~{Nguyen}, A.~R. {Poppe}, R.~{Street}, and M.~{Werner}.
\newblock {Astrophysics with New Horizons: Making the Most of a Generational
  Opportunity}.
\newblock \emph{\pasp}, 130\penalty0 (11):\penalty0 115001, November 2018.
\newblock \doi{10.1088/1538-3873/aadb77}.

\bibitem[{Mather} and {Beichman}(1996)]{Mather1996}
J.~C. {Mather} and C.~A. {Beichman}.
\newblock {EGBIRT and DESIRE; measuring the CIBR at 3 AU.}
\newblock In E.~{Dwek}, editor, \emph{American Institute of Physics Conference
  Series}, volume 348 of \emph{American Institute of Physics Conference
  Series}, pages 271--277, 1996.
\newblock \doi{10.1063/1.49231}.

\bibitem[Matsuura et~al.(2014)Matsuura, Yano, Yonetoku, Funase, Mori,
  Shirasawa, and Group]{Matsuura2014}
Shuji Matsuura, Hajime Yano, Daisuke Yonetoku, Ryu Funase, Osamu Mori, Yoji
  Shirasawa, and The Solar Sail~Working Group.
\newblock Joint planetary and astronomical science with the solar power sail
  spacecraft.
\newblock \emph{TRANSACTIONS OF THE JAPAN SOCIETY FOR AERONAUTICAL AND SPACE
  SCIENCES, AEROSPACE TECHNOLOGY JAPAN}, 12\penalty0 (ists29):\penalty0
  Tr\_1--Tr\_5, 2014.

\bibitem[Bock et~al.(2012)Bock, Beichman, Cooray, Reach, Chary, Werner, and
  Zemcov]{Bock2012}
James Bock, Charles Beichman, Asantha Cooray, William Reach, Ranga-Ram Chary,
  Michael Werner, and Michael Zemcov.
\newblock Astronomical opportunities from the outer solar system.
\newblock \emph{{SPIE} Newsroom}, feb 2012.
\newblock \doi{10.1117/2.1201202.004144}.
\newblock URL \url{https://doi.org/10.1117/2.1201202.004144}.

\bibitem[{Holzer} et~al.(1991){Holzer}, {Mewaldt}, and
  {Neugebauer}]{Holzer1991}
T.~E. {Holzer}, R.~A. {Mewaldt}, and M.~{Neugebauer}.
\newblock {The Interstellar Probe: A Frontier Mission to the Heliospheric
  Boundary and Interstellar Space}.
\newblock \emph{International Cosmic Ray Conference}, 2:\penalty0 535, August
  1991.

\bibitem[Liewer et~al.(2000)Liewer, Mewaldt, Ayon, and Wallace]{Liewer2000}
P.~C. Liewer, R.~A. Mewaldt, J.~A. Ayon, and R.~A. Wallace.
\newblock Nasa's interstellar probe mission.
\newblock \emph{AIP Conference Proceedings}, 504\penalty0 (1):\penalty0
  911--916, 2000.
\newblock \doi{10.1063/1.1302594}.
\newblock URL \url{https://aip.scitation.org/doi/abs/10.1063/1.1302594}.

\bibitem[{McNutt} et~al.(2001){McNutt}, {Andrews}, {McAdams}, {Gold}, {Santo},
  {Ousler}, {Heeres}, {Fraeman}, and {Williams}]{McNutt2001}
R.~L. {McNutt}, Jr., G.~B. {Andrews}, J.~V. {McAdams}, R.~E. {Gold}, A.~G.
  {Santo}, D.~A. {Ousler}, K.~J. {Heeres}, M.~E. {Fraeman}, and B.~D.
  {Williams}.
\newblock {A realistic interstellar probe}.
\newblock In K.~{Scherer}, H.~{Fichtner}, H.~J. {Fahr}, and E.~{Marsch},
  editors, \emph{The Outer Heliosphere: The Next Frontiers}, page 431, 2001.

\bibitem[{Mewaldt} and {Liewer}(2001)]{Mewaldt2001}
R.~A. {Mewaldt} and P.~C. {Liewer}.
\newblock {Scientific Payload for an interstellar probe mission}.
\newblock In K.~{Scherer}, H.~{Fichtner}, H.~J. {Fahr}, and E.~{Marsch},
  editors, \emph{The Outer Heliosphere: The Next Frontiers}, page 451, 2001.

\bibitem[{Fiehler} and {McNutt}(2006)]{Fiehler2006}
D.~I. {Fiehler} and R.~L. {McNutt}.
\newblock {Mission Design for the Innovative Interstellar Explorer Vision
  Mission}.
\newblock \emph{Journal of Spacecraft and Rockets}, 43:\penalty0 1239--1247,
  November 2006.
\newblock \doi{10.2514/1.20995}.

\bibitem[{Wimmer-Schweingruber} et~al.(2009){Wimmer-Schweingruber}, {McNutt},
  {Schwadron}, {Frisch}, {Gruntman}, {Wurz}, and {Valtonen}]{Wimmer2009}
R.~F. {Wimmer-Schweingruber}, R.~{McNutt}, N.~A. {Schwadron}, P.~C. {Frisch},
  M.~{Gruntman}, P.~{Wurz}, and E.~{Valtonen}.
\newblock {Interstellar heliospheric probe/heliospheric boundary explorer
  mission - a mission to the outermost boundaries of the solar system}.
\newblock \emph{Experimental Astronomy}, 24:\penalty0 9--46, May 2009.
\newblock \doi{10.1007/s10686-008-9134-5}.

\bibitem[Stone et~al.(2015)Stone, Alkalai, Friedman, Arora, Arya, Barnes,
  Brashears, Brown, Cauley, Cesarone, Dyson, Garber, Goldsmith, Jemison,
  Johnson, Liewer, Lubin, Maccone, Males, McDonough, Ralph L.~McNutt, Mewaldt,
  Michael, Montgomery, Opher, Provornikova, Rankin, Redfield, Shao, Shotwell,
  Strange, Svitek, Swain, Turyshev, Werner, and Zank]{Stone2015}
Edward Stone, Leon Alkalai, Louis Friedman, Nitin Arora, Manan Arya, Nathan
  Barnes, Travis Brashears, Mike Brown, Paul~Wilson Cauley, Robert~J. Cesarone,
  Freeman Dyson, Darren Garber, Paul Goldsmith, Mae Jemison, Les Johnson,
  Paulett Liewer, Philip Lubin, Claudio Maccone, Jared Males, Kyle McDonough,
  Jr. Ralph L.~McNutt, Richard Mewaldt, Adam Michael, Edward Montgomery, Merav
  Opher, Elena Provornikova, Jamie Rankin, Seth Redfield, Michael Shao, Robert
  Shotwell, Nathan Strange, Thomas Svitek, Mark Swain, Slava Turyshev, Michael
  Werner, and Gary Zank.
\newblock {Science and Enabling Technologies for the Exploration of the
  Interstellar Medium}.
\newblock Technical report, Keck Institute for Space Studies, 04 2015.

\bibitem[{McNutt}(2017)]{McNutt2017}
R.~L. {McNutt}, Jr.
\newblock {Interstellar Probe: First Step to the Stars}.
\newblock \emph{AGU Fall Meeting Abstracts}, December 2017.

\end{thebibliography}

\end{document}